\documentclass[preprint,aps,showpacs]{revtex4}
\usepackage{graphics,rotate}

\begin{document}

\title{New summing algorithm using ensemble computing}
\author{C. D'Helon}
\author{V. Protopopescu}
\email{protopopesva@ornl.gov}
\affiliation{Center for Engineering Science Advanced Research, Computer Science and Mathematics Division,\\ Oak Ridge National Laboratory, Oak Ridge, TN 37831-6355 USA}

\begin{abstract}
We propose an ensemble algorithm, which provides a new approach for evaluating and summing up a set of function samples. The proposed algorithm is not a quantum algorithm, insofar it does not involve quantum entanglement. The query complexity of the algorithm depends only on the scaling of the measurement sensitivity with the number of distinct spin sub-ensembles. From a practical point of view, the proposed algorithm may result in an exponential speedup, compared to known quantum and classical summing algorithms. However in general, this advantage exists only if the total number of function samples is below a threshold value which depends on the measurement sensitivity.
\end{abstract}

\maketitle

\newpage
\section{Introduction}
\label{Introduction}

In this paper, we propose an ensemble algorithm for evaluating and summing an arbitrary function, as an alternative to the quantum algorithms that are currently believed to be the most efficient algorithms available, in terms of query complexity.

The possibility of speeding up the evaluation and summing of a large number of function samples was first noted by Abrams and Williams \cite{Abrams99}, who suggested calculating numerical integrals and stochastic processes using quantum algorithms. Quantum algorithms exploit the inherent parallelism offered by entangled quantum states, to perform certain computational tasks much more efficiently than classical devices, using either \textit{pure} or \textit{pseudopure} quantum states \cite{DiVincenzo00,Vandersypen01} that are tensor products of multiple qubits. The numerical value of an integral is evaluated by employing either the mean estimation algorithm devised by Grover \cite{Grover97b} to calculate the mean of a discrete set of numbers, or by using the quantum counting algorithm proposed by Brassard, Hoyer, and Tapp \cite{Brassard98} to determine the number of elements that fulfill a specified condition. Both of these approaches rely on a generalization of Grover's search algorithm, resulting in a quadratic speedup in comparison with classical randomized (Monte Carlo) algorithms, and an exponential speedup in comparison with classical deterministic algorithms for a single processor.

A systematic comparison of optimal summation of finite sequences and continuous-function integration for deterministic, randomized, and quantum algorithms has been done by Heinrich and Novak \cite{Heinrich2001a,Heinrich2001b}. They have examined the query complexity of quantum integration for different classes of integrand functions, assuming that the critical quantum speedup is obtained by using one of the two quantum summing algorithms mentioned above.

Recently an alternative paradigm for computing has been suggested by Madi, Bruschweiler, and Ernst, which operates on \textit{ensembles} i.e., mixed states of identical spin sytems, using a spin Liouville space formalism \cite{Madi98,Bruschweiler00}. These types of ensemble algorithms are not quantum algorithms, insofar they do not involve entanglement of quantum states. Throughout this paper, "mixed states" describe a statistical ensemble, not individual systems, so that each element of the ensemble performs part of the computation, in the same way as a classical parallel computer.

This new paradigm exploits the parallelism offered by simultaneously acting on linear combinations of many different input states in an ensemble of spins. 
Thus in general, ensemble computing requires an exponentially larger set of memory resources to encode the same number of distinct input states compared to quantum computing with pure states.

While ensemble computing requires more physical resources, it holds the important advantage that it is insensitive to the decoherence time of the spins, which is an outstanding limiting factor for quantum computations involving entangled states. Moreover, ensemble algorithms can be exponentially faster than the equivalent quantum algorithms, for adequate measurement sensitivities, so a trade-off exists between memory and speed capabilities.


We proceed to present a new approach to summing up function samples using an ensemble algorithm, and discuss its query complexity. We only consider the query complexity of the algorithm, i.e., the number of function evaluations that are performed, since the overall computational complexity will depend on the actual function that is being evaluated. At present, the most feasible physical implementation of this summing algorithm would rely on NMR technology, though any physical system of spins can be used in principle. The ideal physical system would allow us to have full control over a very large number of spins, in order to satisfy the large memory requirements. In the Discussion section, we comment on the application of the proposed summing algorithm to evaluating the mean of a continuous function, and as a corollary, on estimating the definite integral of a continuous multi-dimensional function.

\section{Ensemble summing algorithm}
\label{Ensemble summing algorithm}

\subsection{Statement of the Problem}
\label{Statement of the Problem}

Let $f:\{1,2,\dots,N\}\rightarrow [0,1]$ be a real-valued function defined on a discrete set of samples comprised of $N=2^n$ points.  The function $f$ may be known analytically or it may be the result of an explicit or hidden numerical computation. The latter case is known as an \textit{oracle}. We want to evaluate efficiently the sum $S_N$,
\begin{equation}
S_N=\sum_{i=1}^{N}f(i).
\end{equation}
Here efficiency is understood in relation to the query complexity of the algorithm. Indeed, when $N$ is large and the function evaluation is costly in terms of computational complexity, reducing the number of function evaluations is critical.

We assume that the algorithm is to be implemented in a physically realizable system consisting of a finite number of two-valued spins.  To accomodate the $N$ input values, we need  $n$ spins in the input register. The finiteness of the system and the discreteness of the spin states implies that we have to approximate the set of {\it function values}, $\{f(i)\}$, with a set of finite-precision values $\{f_i \in [0,1]\}$ for $i=1,2,\dots,N$. For the sake of simplicity, we shall consistently use the same notation for $f(i)$ and its $k$-digit approximation. Since the meaning of the formulas is explained in the text, there should be no ambiguity. The number of spins $k$ available in the output register will specify the minimal precision, $\delta=2^{-k}$, for these values. Therefore we are actually evaluating the sum $S_{N,k}$,
\begin{equation}
S_{N,k}=\sum_{i=1}^{N}f_i,
\end{equation}
which converges exponentially fast to the sum $S_N$, as we increase the number of spins $k$ in the output register. Thus if we can evaluate $S_{N,k}$ efficiently, we can also evaluate $S_N$ efficiently.

\subsection{Outline of the Algorithm}

The proposed algorithm has three main steps. The first step consists of preparing an ensemble mixture of input states representing the numbers $i=1,2,\dots,N$. In the second step, the function $f$ is applied to the input states, using a single transformation $U_f$ to perform the function evaluation for every input state $i$ at once. This parallel application results in an ensemble mixture which contains all of the values $f_i$ in the output register. Finally, measurement of the output register automatically averages the contributions from the entire ensemble, yielding a signal proportional to the approximate sum, $S_{N,k}$.

\subsection*{Step 1 - Initialization}
We initialize the $n$-spin input register in an equally-weighted mixed state $\rho_{in}^{(n)}$,
\begin{equation}
\rho_{in}^{(n)}=\frac{1}{N}\sum_{i=1}^{N}|i>_n<i|_n,
\label{rho_in}
\end{equation}
which accounts for all $N=2^n$ possible states. The mixed state $\rho_{in}^{(n)}$ is a density operator, which can be represented in spin Liouville space by a density matrix that has non-zero elements only on its diagonal. The off-diagonal elements are all zero, indicating the absence of quantum coherence between any of the states $|i>_n$.

We note again that this mixed state describes a statistical ensemble, thus the initialization is equivalent to assigning each of the input values $i=1,2,\dots,N$, to one of $N$ classical processors in a parallel computer.

For example, in an NMR implementation, the states $|i>_n$ correspond to the eigenstates of the Zeeman Hamiltonian created by a strong external magnetic field \cite{Ernst87}. The ket states, $|i>_n$, can also be written in terms of individual spins,
\begin{equation}
|i>_n=|a_{i1}>\otimes|a_{i2}>\otimes\dots\otimes|a_{in}>
\end{equation}
where $(a_{i1}, a_{i2},\dots,a_{in})\in \{0,1\}$ are the digits of the number $(i-1)$ in binary format.  The bra states, $<i|_n$, are the dual of the ket states.  The state $|0>$ denotes a spin ``up" and the state $|1>$ denotes a spin ``down".
At room temperature, the thermal equilibrium state of an $n$-spin ensemble in an NMR experiment closely approximates the desired initial state $\rho_{in}^{(n)}$ \cite{Gershenfeld97}. The thermal state is equal to the sum of $\rho_{in}^{(n)}$ and a traceless deviation density matrix, with zero off-diagonal terms. The error introduced by using a thermal state instead of the equally-weighted mixed state $\rho_{in}^{(n)}$ is addressed in the Discussion.

We also assume that we have available an output register with $k$ spins, which is capable of encoding the real-numbered values of the series $f_i \in [0,1]$ with precision $\delta=2^{-k}$. All of the states of the output register are initially set to zero, so the state of the entire ensemble (input and output registers) is given by $\rho_{in}^{(n)}\otimes\rho_{out}^{(k)}$,

\begin{equation}
\label{initial_state}
\rho_{in}^{(n)}\otimes\rho_{out}^{(k)}=\frac{1}{N}\sum_{i=1}^{N}|i>_n<i|_n\otimes|0>_k<0|_k.
\end{equation}

\subsection*{Step 2 - Function Evaluation}

The function $f$, analogous to the oracle in Grover's search algorithm, is evaluated by applying a reversible unitary transformation $U_f$. The transformation has no effect on the eigenstates $|i>_n$, but partitions the output register into a set of subensembles, $|f_i>_k$ , defined as the sets of $k$ spins which share the same state. Therefore we have,
\begin{equation}
U_f |i>_n\otimes|0>_k \rightarrow |i>_n\otimes|f_i>_k.
\end{equation}
In a physical implementation, $U_f$ would be the product of a sequence of fundamental unitary transformations for each of the $k$ spins in the output register. The unitary extension of $U_f$ to arbitrary output states uses the bitwise XOR operator $\oplus$:
\begin{equation}
U_f |i>_n\otimes|j>_k \rightarrow |i>_n\otimes|j \oplus f_i>_k.
\end{equation}

The transformation $U_f$ is applied to the system, in order to evaluate the function $f$ \textit{simultaneously} on the linear combination of all sample points $i$ given by the initial state in Eq.(\ref{initial_state}). This is equivalent to evaluating the function $f$ concurrently on $N$ classical processors in a parallel computer.

Since the initial mixed state is a density operator, the action of $U_f$ can be written as
\begin{equation}
\label{final_state}
U_f (\rho_{in}^{(n)}\otimes\rho_{out}^{(k)}) U_f^\dagger = \frac{1}{N}\sum_{i=1}^{N}|i>_n<i|_n\otimes|f_i>_k<f_i|_k \nonumber
\end{equation}

This operation transforms the state of the output register to a mixture that represents all of the approximate function values $f_i$, for $i=1,2,\dots,N$. We remind the reader that the finite-precision values $f_i$ are represented by a binary string of $k$ spins. The states of the output register can be written in terms of individual spins,
\begin{equation}
|f_i>_k=|b_{f_i1}>\otimes|b_{f_i2}>\otimes\dots\otimes|b_{f_ik}>,
\end{equation}
where $(b_{f_i1},b_{f_i2},\dots,b_{f_ik})\in \{0,1\}$ are the digits of the approximate function value $f_i$ in binary format.

We use the following binary encoding scheme to approximate the set of function values $f(i)$ using the $2^k$ states available in the output register:
\begin{eqnarray}
|0_1>\otimes|0_2>\otimes\dots\otimes|0_k> & \leftrightarrow & f(i)\in[0,\delta) \\
|1_1>\otimes|0_2>\otimes\dots\otimes|0_k> & \leftrightarrow & f(i)\in[\delta,2\delta) \\
|0_1>\otimes|1_2>\otimes\dots\otimes|0_k> & \leftrightarrow & f(i)\in[2\delta,3\delta) \\
 & \vdots & \nonumber\\
|1_1>\otimes|1_2>\otimes\dots\otimes|1_k> & \leftrightarrow & f(i)\in[1-\delta,1].
\end{eqnarray}
Alternatively, the approximate function values $f_i$ can be defined directly in terms of the individual spin values, for example,
\begin{equation}
\label{binary}
f_i=\delta\sum_{j=1}^k 2^{j-1} b_{f_ij},
\end{equation}
to set $f_i$ equal to the start of the range intervals given in the encoding scheme above.

Up to this point, the query complexity of the summing algorithm is one i.e., only one function invocation is required. An important, but separate issue that naturally arises is whether $U_f$ can be implemented efficiently. Despite a query complexity of $O(1)$, the computational complexity of the algorithm can be much higher, if the evaluation of the function $f$ is costly. We will not discuss the computational complexity in this paper, since we want to keep $f$ as general as possible, but we note that if $f$ is a classically efficiently computable function, then $U_f$ can be implemented with comparable complexity, as discussed by Nielsen and Chuang \cite{Nielsen}. Moreover, functions that cannot be computed efficiently by classical devices may be rendered tractable in the future by other quantum algorithms.

\subsection*{Step 3 - Measurement}

In the last step of the algorithm, we measure the average value of the output register in the final ensemble given by Eq.(\ref{final_state}). This measurement is an analog process, which performs a single concurrent evaluation, unlike the recursive evaluation on conventional digital computers. The result of the measurement is an ensemble average of all the approximate function values $f_i$,
\begin{equation}
\bar{f_i}=\frac{1}{N} \sum_{i=1}^N f_i, 
\end{equation}
which depends on the distribution of $f_i$ and the precision $\delta=2^{-k}$ of the encoding. Finally the desired sum $S_{N,k}$ can be obtained by multiplying the average value $\bar{f_i}$ and the total number of sample points $N$.

Note that in the statement of the problem we have assumed that the number of function samples is a power of two, $N=2^n$. This ensures that memory resources are employed optimally, by using every possible state of the input register to encode the sample points $i$. However, in general, the number of function samples can be arbitrary, in which case only a subset of the input register states is used to represent the sample points.

In the physical implementation of the algorithm, each of the spins in the output register generates an output signal $\gamma_j$, proportional to the number of spin subensembles that have the $j$-th spin in the state $|1>$. Each signal $\gamma_j$ can be transformed into a fraction $\bar{\gamma}_j \in [0,1]$,
\begin{equation}
\bar{\gamma}_j=\frac{\gamma_j}{\Gamma_j},
\end{equation}
by calibration against the maximum output signal $\Gamma_j$, which is obtained when spin $j$ of the output register is set to $|1>$ for all subensembles. The normalized output signals $\bar{\gamma}_j$ are then multiplied by the corresponding binary weight $2^{j-1}$, cf., Eq.(\ref{binary}), for $j=1,2,\dots k$, to give the ensemble average of the output register,
\begin{equation}
\label{average_f}
\bar{f_i}=\frac{1}{2^k}\sum_{j=1}^k 2^{j-1}\bar{\gamma}_j,
\end{equation}
and hence the sum $S_{N,k}$.

If the measurement sensitivity of the experiment is adequate to distinguish between distinct normalized output signals with a precision equal to or better than $1/N$, then the query complexity remains $O(1)$.

However, as the number of sample points $N$ increases, the sensitivity will eventually become inadequate. At that point, we cannot reliably measure the output signals for each spin in the output register, and significant differences between normalized output signals, differences larger than $1/N$, will not be detectable in a single experimental trial. We note that the error in the ensemble average value $\bar{f_i}$ is given by the weighted sum of the measurement errors for each of the spins in the output register, using the exponentially increasing weights in Eq.(\ref{average_f}). To enhance the measurement sensitivity, the algorithm is repeated a number of times. For example, in an NMR implementation the proposed algorithm will have to be repeated $N^2$ times, taking into account the square-root scaling of the signal-to-noise ratio $S \propto \sqrt{N_e}$ with the number of experimental trials, $N_e$.

\section{Discussion}
\label{Discussion}

The ensemble summing algorithm proposed in this paper uses a radically different approach from previous summing algorithms, which rely either on the extrinsic parallelism of conventional digital computers, or on the intrinsic parallelism of \textit{entangled} quantum states. Instead, the proposed algorithm uses the parallelism of \textit{mixed states} in an ensemble of spins to evaluate a given function once, and then extracts the measurement result by ensemble averaging. We note that the lower bound derived by Nayak and Wu \cite{Nayak98} for the query complexity of quantum algorithms that calculate the mean of a function, does not apply to our ensemble algorithm, since the polynomial method \cite{Beals98} used in their derivation applies to pure states, not mixed states as in the ensemble case. The appropriate lower bound for the ensemble summing algorithm is obtained by considering the query complexity for a classical parallel computer with $N$ processors, which is $O(1)$.

Table \ref{complexity} shows the query complexities of the present ensemble summing algorithm and the ensemble search algorithm proposed by Bruschweiler \cite{Bruschweiler00}. We compare both ensemble algorithms with Grover's search algorithm \cite{Grover97a}, which provides the critical speedup in existing quantum summing algorithms.

\begin{table*}[t!]
\begin{tabular}{|l||c|c|c|c|} \hline
 & Ensemble & Ensemble & Grover's search & Grover's search \\
 & summing & search & (pseudo-pure state) & (pure state) \\ \hline\hline
measurement sensitivity scaling & $1/N$ & $1/N$ & $1/N$ & 1 \\ \hline
no. of NMR experimental trials & $N^2$ & $N^2$ & $N^2$ & 1 \\ \hline
query complexity (single-run) & $O(1)$ & $O(\log N)$ & $O(\sqrt N)$ & $O(\sqrt N)$ \\ \hline
query complexity (overall) & $O(N^2)$ & $O(N^2 \log N)$ & $O(N^2\sqrt N)$ & $O(\sqrt N)$ \\ \hline
\end{tabular}
\caption{Comparison of measurement sensitivity scaling, number of experimental trials required in an NMR implementation, single-run query complexity, and overall query complexity, in the case of inadequate measurement sensitivity i.e., for large $N$.}
\label{complexity}
\end{table*}

The ensemble summing algorithm has an exponential advantage in terms of query complexity, relative to the implementation of the quantum summing algorithm using Grover's search algorithm with pseudopure states in NMR, as the two algorithms have the same scaling for the measurement sensitivity. Similarly, the query complexity of the ensemble search algorithm is exponentially smaller than that required for Grover's search algorithm using pseudopure states in NMR.

However, a comparison with the (theoretical) implementation of Grover's search algorithm using pure states shows that both the ensemble summing algorithm and the ensemble search algorithm will be more efficient only for a total number of samples below a threshold value determined by the measurement sensitivity. Bruschweiler \cite{Bruschweiler00} estimates that the ensemble search algorithm is more efficient for databases of size $N$ which fulfill the condition
\begin{equation}
N \sqrt N \log_2 N < S^2,
\end{equation}
where $S$ is the signal-to-noise ratio of measurements in an NMR implementation.

The same reasoning leads to the conclusion that the ensemble summing algorithm is more efficient than the quantum summing algorithm using Grover's search algorithm with pure states, for a number of function samples $N$ given by
\begin{equation}
N \sqrt N < S^2,
\end{equation}
with respect to the signal-to-noise ratio $S$. The best available signal-to-noise ratio in present NMR technology is $S \approx 10^4$, which results in an efficiency threshold value $N_{max} \approx 2\times10^5$. However for values of $N > N_{max}$, the query complexity of the ensemble summing algorithm is $O(N^2)$, which is greater than the query complexity for quantum and classical summing algorithms.

A different type of error occurs in the initialization step of the algorithm, if a room-temperature thermal state is used instead of the initial state $\rho_{in}^{(n)}$ in Eq. (\ref{rho_in}). For an NMR implementation, Gershenfeld and Chuang \cite{Gershenfeld97} give the thermal state for $n$ spins in the form
\begin{equation}
\rho_{th}^{(n)}=\frac{1}{N}\sum_{i=1}^{N}|i>_n<i|_n + \frac{\alpha}{N}\sum_{i=1}^{N}\chi_i |i>_n<i|_n,
\end{equation}
where $\alpha=\frac{\hbar \omega}{2 k_B T}$ ($\approx 10^{-6}$ at room temperature) is the Boltzmann factor of the deviation density matrix, and the coefficients $\chi_i \in [-n,n]$ represent the net integer sum of various spin-up and spin-down combinations of the $n$ spins.

The effect of using a thermal state instead of the equally-weighted mixed state $\rho_{in}^{(n)}$ becomes present in the measurement step of the algorithm, which results in an ensemble average
\begin{equation}
\bar{f_i}^\prime=\frac{1}{N} \sum_{i=1}^N f_i + \frac{\alpha}{N} \sum_{i=1}^N \chi_i f_i. 
\end{equation}
The error term in this average imposes an additional limit on the accuracy of the algorithm. In the worst case, the difference between the measured and desired values is bounded by
\begin{equation}
|\bar{f_i}^\prime - \bar{f_i}| \leq \frac{n\alpha}{2},
\end{equation}
since $|\chi_i|\leq n$ and $f_i \leq 1$.
In general, the error is much smaller than $\frac{n\alpha}{2}$, as the deviation matrix is traceless, i.e., $\sum \chi_i = 0$. The Boltzmann factor $\alpha$ can also be reduced by either raising the temperature $T$, or by choosing a lower average resonant frequency $\omega$ for the spins.

The ensemble summing algorithm presented in this paper can be applied to estimating the mean and/or the definite integral of a multi-dimensional function. The validity of the algorithm holds for rather general classes of functions, ranging from continuous to the Lebesgue measurable and integrable classes, $L^p, 1 \leq q < \infty$.  The former is based on the convergence of the Riemann sums to the Riemann integral of "ordinary" functions, while the latter is based on the density of simple (Boolean) functions in $L^p, 1 \leq q < \infty$ \cite {DS}.  Unfortunately, these results simply state that in the limit of infinite number of terms, the approximating sums coincide with the desired integrals.  An evaluation of the error made when using a finite number of terms in the sum is impossible, in general.  To estimate this error, one has to resort to the specific properties of the approximated (sub-classes of) functions. For instance, for any Lipschitz function $f$ with Lipschitz constant $L$, integrated over the finite interval $[a, b]$, the error, $E_N$, between the integral of $f$ and the Riemann sum with $N$ terms evaluated at equidistant points is bounded by $(b - a)L/N$.  This error can now be combined with the error made when estimating the discrete sum with $N$ terms (see Section \ref{Statement of the Problem}) and an efficient algorithm can then be devised for the estimation of the integral. For general functions, the expression of the error as a function of $N$ is unknown, although it is known that $lim_{N\rightarrow\infty} E_N = 0$.  However, if $E_N$ decreases with $N$ in a much slower fashion, say like $E_N\approx(ln N)^{-1}$, this would translate into a significant increase of the number of terms in the sum and therefore an increased complexity, to achieve a given overall precision for the integral. The relationship between various functional classes and their approximants by Boolean functions is an active research topic that addresses such notions as the complexity, capacity, and entropy of a function, which go beyond the scope of the present paper.  The interested reader is referred to Refs. \cite {Heinrich2001a,Heinrich2001b,L,TWW,V}.

We conclude by pointing out the two main advantages of the proposed ensemble summing algorithm. First, there is no need to maintain quantum coherences for ensemble algorithms, so they are easier to implement than their quantum counterparts. Indeed, the scaling of the resources required to maintain entanglement in pure state-based algorithms for the duration of the computation remains an open question, and could have a potentially large negative effect on the exact threshold value $N_{max}$ of the number of function samples.

Second, for a restricted number of function samples, $N \ll N_{max}$, which is determined by the measurement sensitivity, ensemble algorithms may give an exponential speedup over all known quantum and classical summing algorithms. In this regime, the proposed ensemble summing algorithm requires only a single invocation of the function $f$.

\section* {Acknowledgments}
\label {Acknowledgments}
{This research was partly supported by the U.S. Department of Energy, Office of Basic Energy Sciences. The Oak Ridge National Laboratory is managed for the U.S. DOE by UT-Battelle, LLC, under contract No. DE-AC05-00OR22725.}


\begin{thebibliography}{99}
\bibitem{Abrams99}Abrams D S and Williams C P 1999 \textit{Preprint} quant-ph/9908083
\bibitem{Grover97b}Grover L K 1997 \textit{Preprint} quant-ph/9711043
\bibitem{Brassard98}Brassard G, Hoyer P and Tapp A 1998 \textit{Lect. Notes in Comp. Science} \textbf{1443} 820
\bibitem{Heinrich2001a}Heinrich S and Novak E 2001 \textit{Preprint} quant-ph/0105114
\bibitem{Heinrich2001b}Heinrich S 2001 \textit{Preprint} quant-ph/0112152
\bibitem{DiVincenzo00}DiVincenzo D P 2000 \textit{Preprint} quant-ph/0002077
\bibitem{Vandersypen01}Vandersypen L M K,  Steffen M, Breyta G, Yannoni C S, Sherwood M H and Chuang I L 2001 Nature \textbf{414} 883
\bibitem{Shor94}Shor P 1994 \textit{Proceedings of the 35th Annual Symposium on Foundations of Computer Science} ed. Goldwasser S (IEEE Computer Society: Los Alamitos, CA)
\bibitem{Grover97a}Grover L K 1997 \textit{Phys. Rev. Lett.} \textbf{79} 325
\bibitem{Deutsch}Deutsch D and Jozsa R 1992 \textit{Proc. Roy. Soc. London A} \textbf{439} 553
\bibitem{Simon}Simon D R 1997 \textit{SIAM J. Comput.} \textbf{26}(5) 1474
\bibitem{Madi98}Madi Z L, Bruschweiler R and Ernst R R 1998 \textit{J. Chem. Phys.} \textbf{109} 10603
\bibitem{Bruschweiler00}Bruschweiler R 2000 \textit{Phys. Rev. Lett.} \textbf{85} 4815
\bibitem{Ernst87}Ernst R R, Bodenhausen G and Wokaun A 1987 \textit{Principles of Nuclear Magnetic Resonance in One and Two Dimensions} (Clarendon: Oxford)
\bibitem{Gershenfeld97}Gershenfeld N A and Chuang I L 1997 \textit{Science} \textbf{275} 350
\bibitem{Nielsen}Nielsen M A and Chuang I L 2000 \textit{Quantum Computation and Quantum Information} (University Press: Cambridge)
\bibitem{Nayak98}Nayak A and Wu F 1998 \textit{Preprint} quant-ph/9804066
\bibitem{Beals98}Beals R, Buhrman H, Cleve R, Mosca M and de Wolf R 1998 \textit{Preprint} quant-ph/9802049
\bibitem{DS}Dunford N and Schwartz J T 1958 \textit{Linear Operators} (Interscience Publishers: New York) vol. 1
\bibitem{L}Lorentz G G 1986 \textit{Approximation of Functions} (Chelsea Publishing Company: New York)
\bibitem{TWW}Traub J F, Wasilkowski G W and Wozniakowski H 1988 \textit{Information-Based Complexity} (Academic Press: Boston)
\bibitem{V}Vapnik V N 1995 \textit{The Nature of Statistical Learning} (Springer: New York)
\end{thebibliography}
\end{document}